# Elastic properties of Fe−C and Fe−N martensites

SOUISSI Maaouia [a, b] and NUMAKURA Hiroshi [a, b] *

[a] Department of Materials Science, Osaka Prefecture University, Naka-ku, Sakai 599-8531, Japan

[b] JST-CREST, Gobancho 7, Chiyoda-ku, Tokyo 102-0076, Japan

*Corresponding author. E-mail: numakura@mtr.osakafu-u.ac.jp

| | |
|---|---|
| Postal address: | Department of Materials Science, Graduate School of Engineering, Osaka Prefecture University, Gakuen-cho 1-1, Naka-ku, Sakai 599-8531, Japan |
| Telephone: | +81 72 254 9310 |
| Fax: | +81 72 254 9912 |

SYNOPSIS

Single-crystal elastic constants of bcc iron and bct Fe–C and Fe–N alloys (martensites) have been evaluated by ab initio calculations based on the density-functional theory. The energy of a strained crystal has been computed using the supercell method at several values of the strain intensity, and the stiffness coefficient has been determined from the slope of the energy versus square-of-strain relation. Some of the third-order elastic constants have also been evaluated. The absolute magnitudes of the calculated values for bcc iron are in fair agreement with experiment, including the third-order constants, although the computed elastic anisotropy is much weaker than measured. The tetragonally distorted dilute Fe–C and Fe–N alloys exhibit lower stiffness than bcc iron, particularly in the tensor component $C_{33}$, while the elastic anisotropy is virtually the same. Average values of elastic moduli for polycrystalline aggregates are also computed. Young's modulus and the rigidity modulus, as well as the bulk modulus, are decreased by about 10 % by the addition of C or N to 3.7 atomic per cent, which agrees with the experimental data for Fe–C martensite.

# 1. Introduction

Iron–carbon martensite is the most important constituent in providing high strength to steels. A fully martensitic structure produces the maximum hardness in conventional carbon steel, and the martensite is the key component also in modern advanced products such as DP (dual phase), TRIP (transformation-induced plasticity), and Q&P (quench and partitioning) steels[1-3]. The strength, or hardness, of martensite is believed to be caused by solid-solution strengthening, high density of dislocations, as well as its complex microstructure resulting from the characteristic cubic-to-tetragonal transformation. The elastic stiffness of the martensite is also of interest, first for its own sake as one of the mechanical properties, and second because elastic misfit is an important factor controlling the microstructure formation, in addition to lattice misfit. These effects have been studied extensively for diffusion-controlled phase transformations[4-6]. Recently, simulation studies of martensitic transformations focussing on effects of elastic properties have also been made[5-10].

The elastic moduli of iron-carbon martensite were reported by several researchers[11-15]. Schmidtmann et al.[11] measured Young's modulus of quenched carbon steels by the resonance vibration method and found that it decreases with increasing the C content. Speich et al.[13] obtained similar results for both Young's modulus and shear modulus of quenched Fe–C alloys. However, the report by Dey et al.[12] was contradictory: their sound velocity measurements on Fe–Ni–C martensites indicated that Young's modulus and shear modulus increase with the C content. This discrepancy seems to have been resolved by the analysis by Speich and Leslie[14] that estimated the effects of Ni content and retained austenites; it was then established that the hard martensite is in fact elastically softer than ferrite, i.e., bcc iron. More recently, single-crystal elastic constants of the martensite phase in ferrite–martensite two-phase mixtures were determined, together with those of the ferrite phase, by resonant ultrasound spectroscopy[15]. The elastic moduli of polycrystals derived from them are lower than those of iron, in accordance with the conclusion of Speich and Leslie.

Young's modulus and shear modulus of polycrystalline materials provide sufficient material data in practice, but more detailed information, i.e., single-crystal elastic constants, is desired in, for example, understanding microstructure formation, where the mismatch in the elastic stiffness and its anisotropy may play a critical role. While a single-phase martensite material can readily be prepared, producing its single crystal is prohibitively difficult. It is the purpose of this work to theoretically evaluate the single-crystal elastic constants of iron–carbon and iron–nitrogen martensites by means of ab initio calculations. As mentioned above, the single-crystal elastic constants of Fe–C martensite were evaluated by sophisticated measurements and analysis of ferrite–martensite two-phase mixtures[15], but only at a single C content. A difficulty in experiment is, either of single-phase or in two-phase mixtures, martensites always involve high density of crystal defects and are accompanied with internal stresses arising from the structural transformation; experimentally determined values of elastic modulus may be influenced by these defects and stresses. From the theoretical side a simulation study has recently been made on the elastic constants of Fe–C martensite of C contents up to 2.2 atomic per cent[16], but using semi-empirical interatomic potentials. In this study we have determined the elastic constants on the basis of the density-functional theory (DFT) using the supercell technique as a function of the solute content up to 3.7 atomic per cent. It is also of interest to find similarities or dissimilarities between Fe–C and Fe–N alloys in elastic properties. To the best of our knowledge, no reports on the elasticity of Fe–N martensite are found in the literature[1].

---

[1] In this paper we use the term 'martensite' to indicate primarily the dilute Fe–C and Fe–N alloy phases of the bct structure. References to microstructures with the same term can be distinguished from the context.

## 2. Method

### 2.1. Determination of elastic constants

The elastic stiffness constants of bcc iron and tetragonal martensites have been determined from the relation between the elastic strain energy and the magnitude of the strain, by imposing a strain tensor that corresponds to a deformation pattern characteristic of the crystal symmetry. With application of strain

$$\begin{pmatrix} e_1 & \frac{1}{2}e_6 & \frac{1}{2}e_5 \\ \frac{1}{2}e_6 & e_2 & \frac{1}{2}e_4 \\ \frac{1}{2}e_5 & \frac{1}{2}e_4 & e_3 \end{pmatrix}, \quad (1)$$

the increase in the total energy, or the elastic strain energy, $\Delta E$, of a crystal of volume $V$ is written as

$$\frac{\Delta E(e_1, e_2, ..., e_6)}{V} = \frac{1}{2!}\sum_i \sum_j C_{ij} e_i e_j + \frac{1}{3!}\sum_i \sum_j \sum_k C_{ijk} e_i e_j e_k + O[e^4], \quad (2)$$

where $C_{ij}$ and $C_{ijk}$ are the second-order and the third-order elastic stiffness constants, respectively[17], and $O[e^4]$ indicates the terms of the forth and higher order in the strain magnitude, $e$. The sums are taken over the indices of Voigt notation from 1 to 6.

For crystals of cubic symmetry, the above expression is reduced to

$$\frac{\Delta E}{V} = \frac{1}{2}C_{11}(e_1^2 + e_2^2 + e_3^2) + C_{12}(e_2 e_3 + e_3 e_1 + e_1 e_2) + \frac{1}{2}C_{44}(e_4^2 + e_5^2 + e_6^2)$$
$$+ \frac{1}{6}C_{111}(e_1^3 + e_2^3 + e_3^3) + \frac{1}{2}C_{112}(e_1^2 e_2 + e_1^2 e_3 + e_2^2 e_3 + e_2^2 e_1 + e_3^2 e_1 + e_3^2 e_2)$$
$$+ C_{123} e_1 e_2 e_3 + \frac{1}{2}C_{144}(e_1 e_4^2 + e_2 e_5^2 + e_3 e_6^2)$$
$$+ \frac{1}{2}C_{155}(e_2 e_4^2 + e_3 e_4^2 + e_1 e_5^2 + e_3 e_5^2 + e_1 e_6^2 + e_2 e_6^2) + C_{456} e_4 e_5 e_6. \quad (3)$$

**Table 1** presents the strain tensors employed and the coefficients in the second-order and third-order terms of the magnitude of the strain, $x$. The strains produce deformations corresponding to the three symmetrized second-order elastic constants of cubic crystals, viz., the bulk modulus $B = (C_{11} + 2C_{12})/3$, the $\{1\bar{1}0\}\langle 110\rangle$ shear modulus $C' = (C_{11} - C_{12})/2$, and the $\{100\}\langle 010\rangle$ shear modulus $C_{44}$. These constants are given essentially by the second-order coefficients, which are listed in the third column of the table. In this paper these deformation patterns are referred to as 1, 2a, and 2b, respectively. Since the first is asymmetric with respect to the sign of $x$, it has a third-order term in the elastic energy. Its coefficient gives the third-order bulk modulus, $B^{(3)}$, which is defined by the formula

$$\frac{\Delta E}{V} = \frac{1}{2!} B \left(\frac{\Delta V}{V}\right)^2 + \frac{1}{3!} B^{(3)} \left(\frac{\Delta V}{V}\right)^3. \qquad (4)$$

In deformation 1, $\Delta V/V$ is equal to $3x$, and the relation

$$B^{(3)} = \frac{1}{9}(C_{111} + 6C_{112} + 2C_{123}) \qquad (5)$$

is readily found by letting $e_1 = e_2 = e_3 = x$ and $e_4 = e_5 = e_6 = 0$ in Eq. (3) and comparing the final form with Eq. (4). The third-order coefficient in the table is 9/2 times $B^{(3)}$. In the other two strain patterns are involved no third-order terms.

For tetragonal crystals, there are six independent second-order elastic stiffness constants, $C_{11}, C_{12}, C_{13}, C_{33}, C_{44}$, and $C_{66}$, and twelve third-order constants. The expression of the strain energy is

$$\begin{aligned}
\frac{\Delta E}{V} &= \frac{1}{2}C_{11}(e_1^2 + e_2^2) + \frac{1}{2}C_{33}e_3^2 + C_{12}e_1e_2 + C_{13}(e_1e_3 + e_2e_3) \\
&+ \frac{1}{2}C_{44}(e_4^2 + e_5^2) + \frac{1}{2}C_{66}e_6^2 \\
&+ \frac{1}{6}C_{111}(e_1^3 + e_2^3) + \frac{1}{6}C_{333}e_3^3 + \frac{1}{2}C_{112}(e_1^2e_2 + e_2^2e_1) + \frac{1}{2}C_{113}(e_1^2e_3 + e_2^2e_3) \\
&+ \frac{1}{2}C_{133}(e_1e_3^2 + e_2e_3^2) + \frac{1}{2}C_{144}(e_1e_4^2 + e_2e_5^2) + \frac{1}{2}C_{155}(e_1e_5^2 + e_2e_4^2) \\
&+ C_{123}e_1e_2e_3 + C_{456}e_4e_5e_6 \\
&+ \frac{1}{2}C_{166}(e_1e_6^2 + e_2e_6^2) + \frac{1}{2}C_{344}(e_3e_4^2 + e_3e_5^2) + \frac{1}{2}C_{366}e_3e_6^2.
\end{aligned} \quad (6)$$

We evaluated the six constants by employing the six strain patterns proposed by Mehl et al.[18], which are based on symmetrized elastic constants[19]. The strain tensors and the corresponding second-order and third-order coefficients are listed in **Table 2**. They are classified into two groups, 1a to 1c, and 2a to 2c. For the first three the strain energy is asymmetrical with the sign of $x$, so that they contain terms with third-order elastic constants[^2]. On the other hand, for the second three, which are $\{110\}\langle 1\bar{1}0\rangle$, $\{001\}\langle 110\rangle$, and $\{100\}\langle 010\rangle$ shear deformations, it is symmetrical about $x$ and thus involves no third-order terms.

## 2.2. Ab initio calculation

The calculations were performed by Vienna Ab initio Simulation Package (VASP, version 4.8)[20], using the projector augmented-wave (PAW) method[21] and the general gradient approximation (GGA) to DFT based on Perdew-Burke-Ernzerhof (PBE) functional[22]. The cut-off energy was chosen to be 520 eV[^3], which is 1.3 times the recommended value in the default setting.

We chose the conventional bcc unit cell as the supercell for computations of pure iron.

---

[^2]: In deformation 1a the third-order coefficient involves second-order constants because of the presence of $x^2$ term in $e_3$.

[^3]: 1 eV = $1.602 \times 10^{-19}$ J.

To model Fe–C and Fe–N martensites, a supercell consisting of 3 × 3 × 3 bcc unit cells (54 Fe atoms) was used, with one C or N atom placed in an octahedral interstitial site, or two of the same solute atoms. The solute concentration, $y$, which we define as the number of solute atoms per host atom, is 1/54 (= 0.0185) and 2/54 (= 0.0370), respectively. In the latter case, with the first solute atom at the centre of the supercell, the second solute atom was placed at the corner, which is an octahedral site in the same orientation as the first, and is at a distance $(3/2)\sqrt{3}\,a$, where $a$ is the lattice parameter of bcc iron. This configuration is reported to be the most favourable among all possible configurations of two C atoms in the 3 × 3 × 3 supercell[23].

Brillouin zone integrations were performed using a Monkhorst-Pack k-point mesh of 21 × 21 × 21 for the bcc unit cell of iron, or of 8 × 8 × 8 for the supercell of 3 × 3 × 3 unit cells. The structure and energy were fully optimized until the total energy converges within $10^{-5}$ eV and the magnitudes of the Hellmann-Feynman forces on all atomic sites become less than $10^{-1}$ eV nm$^{-1}$. All computations were done under a spin-polarized condition.

The lattice parameters and the axial ratio, $c/a$, of fully optimized supercells are shown in **Fig. 1**, together with experimental data[24]. The absolute values of the computed lattice parameters are appreciably smaller than the experimental ones, which is a known problem[4], but the axial ratio is in good agreement with experiment for both Fe–C and Fe–N alloys.

## 3. Results and discussion

### 3.1. Pure iron

For each pattern of deformation the total energy was computed at several different values of the strain parameter $x$, ranging from $1 \times 10^{-3}$ to $6 \times 10^{-2}$. The strain energy $\Delta E$ per unit cell is plotted against the square of the strain parameter as a double logarithmic diagram in **Fig. 2**. The data points would form a straight line parallel to the diagonal if the strain energy is proportional to $x^2$. It occurs in the range of $x^2$ around $10^{-5}$, $5 \times 10^{-5}$, and $10^{-4}$ for deformations 1, 2a, and 2b, respectively. In the latter two the linear behaviour extends up to the largest values of $x^2$ examined. On the other hand, it breaks down at small strains, which must be due to limited accuracy of numerical computation. After examining the whole results for Fe–C and Fe–N alloys, some of which are shown later (Fig. 4), we found that $\Delta E$ is satisfactorily linear in $x^2$ when its magnitude is around $10^{-3}$ eV (per unit cell). On the basis of this finding, we set a practical criterion for selecting data points to be used for determining elastic constants in terms of the magnitude of the strain energy, not of strain: $\Delta E$ per unit cell be of the order of $10^{-3}$ eV. For 2a and 2b in Fig. 2, several points that meet this criterion are selected and a linear function is fitted to the $\Delta E$ versus $x^2$ relation by the method of least squares. The results are shown by thick solid lines. Their slope, i.e., the proportionality constant, gives $C_{11} - C_{12}$ and $C_{44} / 2$, respectively (see Table 1).

---

[4] The calculated lattice parameter of pure iron, 283.4 pm, is much smaller than the experimental value at room temperature, 286.6 pm. Extrapolation to absolute zero temperature, 286.05 pm using the thermal expansion data [34], does not much ameliorate the problem.

In deformation 1, the energy–strain relation is inherently asymmetric with respect to the sign of $x$ (uniform dilatation or compression). At large strains the data points diverge to two separate curves, which is the expected behaviour under 'finite elastic deformation'[25]. At small strains the positive and negative branches also begin to deviate from each other and tend to fall downwards from the linear relation, although it may not be discernible at the resolution of this figure. We have chosen several data points that do not suffer from these problems, in the range of energy of the order of $10^{-3}$ eV, and determined the second- and the third-order coefficients by fitting a cubic polynomial (without a constant term nor the first order term), as shown in **Fig. 3**. The second- and the third-order bulk moduli, $B$ and $B^{(3)}$, obtained from them are shown in the first row of **Table 3**, with uncertainty margins originated from the least-squares fitting. The value of the third-order bulk modulus, $-1 \times 10^{12}$ Pa, is found to be close to that evaluated from tensile large-deformation experiments on single-crystal whiskers[26]. For readers' convenience, numerical values of the elastic compliance constants are given in **Appendix A**.

The calculated values of the second-order elastic constants are in general larger in magnitude than the experimental values at 4.2 K[27], $B$ by 9 % and $C'$ by 25 %, for example. Disagreement in the opposite trend is found in the shear modulus $C_{44}$: the calculated value is smaller by 20 % than experiment. They result in the shear anisotropy factor, $A$, defined as the ratio of $C_{44}$ to $C'$, considerably smaller than the experimental one, 1.5 against 2.3 or 2.4. In the lower part of Table 3, results of theoretical calculations found in the literature are shown for comparison. None of them reproduces the strong elastic anisotropy of bcc iron.

### 3.2. Fe-C and Fe-N martensites

Similarly to the case of bcc iron, strain energies have been calculated as a function of the strain parameter $x$ for the six deformation patterns of Table 2. The results for the Fe–C alloy of $y_C = 0.0185$, i.e., of the supercell $Fe_{54}C_1$, are shown in **Figs. 4** and **5** for those of the first type, 1a to 1c (extension / contraction), and in **Fig. 6** for those of the second type, 2a to 2c (shear deformation).

In Fig. 4, failure of the linear relation at small strains is evident, in addition to the natural departure of the extension and contraction curves from each other at large strains. The data are replotted in Fig. 5 against $x$, and selected data points of $\Delta E$ of the order of $10^{-3}$ eV, indicated by the markers for each set, are analysed by fitting a cubic polynomial to determine the second-order and the third-order coefficients. The results of the shear deformations are analysed in a simpler manner by fitting a linear function to $\Delta E$ versus $x^2$ data (Fig. 6). The elastic stiffness constants have been obtained through the same procedure for $Fe_{54}C_2$, $Fe_{54}N_1$, and $Fe_{54}N_2$, and the results are summarized in **Table 4** (the second-order constants) and in **Table 5** (the third-order constants). In the former, the stiffness constants of 'hardened plain carbon steel' reported by Kim and Johnson[15] are shown for comparison. No information on higher-order elastic constants of martensites is found in the literature.

### 3.3. Effect of solute content

The second-order stiffness constants are displayed in **Fig. 7** as a function of the solute concentration, $y$. The second-order bulk modulus is obtained as the reciprocal of the volume compressibility[5], if we ignore the third- and higher-order constants. For tetragonal crystals, it is given from the stiffness constants by the following formula:

$$B = \frac{(C_{11}+C_{12})C_{33} - 2C_{13}^2}{C_{11}+C_{12}+2C_{33}-4C_{13}}. \tag{7}$$

Error bars in the present data are appreciable only in $B$, as it is computed from four other numbers, each of which has uncertainties of 1 to 2 %. For the other quantities, error bars are of similar lengths to the size of the symbols in the figures, and are omitted for clarity.

The general trends in Fe–C alloys, Fig. 7 (a), and those in Fe–N alloys, Fig. 7 (b), are essentially the same: some of the elastic constants decrease with increasing the solute content, while some others, $C_{66}$ and $C_{44}$, exhibit recovery at $y = 0.037$. Upon changing the crystal structure from cubic to tetragonal by the addition of the interstitial solute, each of the three elastic constants divides itself into two:

$C_{11} \rightarrow C_{11}$ and $C_{33}$,

$C_{12} \rightarrow C_{12}$ and $C_{13}$,

and

---

[5] The volume compressibility, $K$, is expressed in terms of the elastic compliance constants $S_{ij}$ as $K = S_{11} + S_{22} + S_{33} + 2(S_{23} + S_{31} + S_{12})$ regardless of crystal symmetry.

$C_{44} \to C_{44}$ and $C_{66}$.

It turns out that significant departure occurs only between $C_{11}$ and $C_{33}$: there is notable lowering of $C_{33}$ in both Fe–C and Fe–N alloys, amounting to a relative change of −18 % (Fe–C) or −17 % (Fe–N) at $y = 0.037$. This may be understood simply as weakening of atomic bonding because of increased interatomic distances in the direction of the tetragonal axis. The component $C_{11}$ also decreases, yet at a lower rate than for $C_{33}$, even though the dimensions of the crystal normal to the tetragonal axis are virtually unchanged: the relative change at $y = 0.037$ is −6 %. About a half of this lowering can be attributed to the increase in the volume of the crystal, of 3.6 %, in the alloys at this concentration; increase in the volume leads to decrease in elastic stiffness constants per se, as the elastic energy is defined as the energy per unit volume.

Concerning the shear stiffness constants, the two shear moduli of bcc iron split in the tetragonal martensite as

$$C' \to (C_{11} - C_{12})/2 \text{ and } (C_{11} + C_{33} - 2C_{13})/4,$$

and

$$C_{44} \to C_{66} \text{ and } C_{44},$$

while the bulk modulus $B$ remains unique. The relative change in $B$ is −10 % at $y_C = 0.037$, but those in the shear stiffness vary: −5.4 % in $(C_{11} - C_{12})/2$, −14.7 % in $(C_{11} + C_{33} - 2C_{13})/4$, −2.3 % in $C_{66}$, and −11.2 % in $C_{44}$. We define two shear anisotropy factors, $A_{44}$ and $A_{66}$, as follows:

$$A_{44} \equiv \frac{C_{44}}{(C_{11} + C_{33} - 2C_{13})/4}, \tag{8}$$

$$A_{66} \equiv \frac{C_{66}}{(C_{11} - C_{12})/2}. \tag{9}$$

Their values are shown in the upper part of Figs. 7 (a) and 7 (b). Both of them are about 1.5 and are almost unchanged in the concentration range studied. The shear anisotropy of martensite is thus suggested to remain similar to that of bcc iron.

The single-crystal elastic constants of iron and Fe–C martensite, of 2.4 % C, reported by Kim and Johnson[15] are shown together in Fig. 7 (a). Their values for bcc iron are all in good agreement with the established data[27], but those of the martensite are at variance with the results of the present study: their $C_{11}$ and $C_{33}$ are equal to each other and are much larger than $C_{11}$ of unalloyed iron, and the two shear moduli derived from $C'$ exhibit a similar trend, while the other two shear moduli, $C_{44}$ and $C_{66}$, decrease together, both much more significantly than do the present results. As shown later in Fig. 12, the magnitudes of the bulk modulus, Young's modulus, and rigidity modulus of polycrystalline materials calculated by averaging the values of Kim and Johnson appear reasonable in reference to the earlier experimental data, but this apparent agreement, or coincidence, seems to result from the compensating variations of the elements, which are rather irregular and do not seem to have a reasonable explanation. In particular, in their report the four shear moduli, $C_{44}$, $C_{66}$, $(C_{11}-C_{12})/2$, and $(C_{11}+C_{33}-2C_{13})/4$, all fall very close to each other, and therefore the martensite is concluded to be elastically highly isotropic. On the other hand, the prediction of the present calculations is that, as remarked above, the elastic anisotropy of the martensites is not very different from that of bcc iron, putting aside the failure in reproducing the strong anisotropy of the latter. We understand that the method of analysis employed by Kim and Johnson is very refined and well established, but it is a not a simple task either to determine single-crystal elastic constants of low-symmetry crystals from such complicated samples in the microstructure.

## 3.4. Anisotropy of practical elastic moduli

Variations of Young's modulus and rigidity modulus with orientation are of practical interest. We show them as three-dimensional (3D) surfaces and as cross-sectional views in **Figs. 8** and **9** (Young's modulus $E$) and in **Figs. 10** and **11** (rigidity modulus $G$) for pure iron and Fe–C and Fe–N martensites of $y = 0.037$. They were computed from the following standard formulae[35]. For cubic crystals, the reciprocal Young's modulus and that of rigidity modulus are given as

$$E^{-1} = S_{11} - \left[2(S_{11} - S_{12}) - S_{44}\right]\Gamma , \qquad (10)$$

$$G^{-1} = S_{44} + 2\left[2(S_{11} - S_{12}) - S_{44}\right]\Gamma , \qquad (11)$$

where

$$\Gamma = \gamma_2^2\gamma_3^2 + \gamma_3^2\gamma_1^2 + \gamma_1^2\gamma_2^2 \qquad (12)$$

is the so-called orientation factor, with $\gamma_1$, $\gamma_2$, $\gamma_3$ being the direction cosines between the stress axis and the three principal crystal axes. For tetragonal crystals, they are given as

$$E^{-1} = S_{11}(\gamma_1^4 + \gamma_2^4) + S_{33}\gamma_3^4 + (2S_{12} + S_{66})\gamma_1^2\gamma_2^2 + (2S_{13} + S_{44})\gamma_3^2(1 - \gamma_3^2), \qquad (13)$$

$$G^{-1} = \frac{1}{2}(S_{44} + S_{66})(\gamma_1^4 + \gamma_2^4) + S_{44}\gamma_3^4 + \left[4(S_{11} - S_{12}) + (S_{44} - S_{66})\right]\gamma_1^2\gamma_2^2 \\ + \left[2(S_{11} + S_{33} - 2S_{13}) + \frac{1}{2}(S_{66} - S_{44})\right]\gamma_3^2(1 - \gamma_3^2). \qquad (14)$$

The 3D surfaces of Young's modulus (Fig. 8) are bulged out in $\langle 111 \rangle$ directions, since the shear anisotropy factor is greater than unity in every case. For the rigidity modulus (Fig. 10) the reverse is true; the 3D surfaces are protruded in $\langle 100 \rangle$ directions. Again it is evident in these illustrations and diagrams that the anisotropy is much less accentuated in the present results than in real bcc iron. It does not seem, therefore, worth examining the directional variations in further detail.

### 3.5. Elastic moduli of polycrystalline aggregates

Another result of practical utility is the elastic moduli of polycrystalline aggregates. Young's modulus, rigidity modulus and Poisson's ratio have been computed by the methods of Voigt (uniform stress assumption), Reuss (uniform strain assumption), and Hill (or Voigt-Reuss-Hill, which is a simple arithmetic average of the former two)[36,37], and are displayed as a function of the solute content in **Fig. 12**. The bulk modulus is also shown for comparison.

While no significant variation occurs in Poisson's ratio with the solute content (it stays in the range between 0.30 and 0.31), the three moduli are found to be lowered by the addition of the solute at similar rates: the decrease is about 10 % at $y = 0.037$. The experimental values for Fe–C martensite by Schmidtmann et al.[11], Speich et al.[13], and Kim and Johnson[15] are shown in Fig. 12 (a). The present results on Fe–C alloys agree fairly well with experiment, not only in the dependencies on the C content but also in the absolute magnitudes. They seem to support the conjecture of Speich et al. that the elastic stiffness of the martensite is lower than that of bcc iron, and provide an explanation that the lower stiffness is primarily due to the significant decrease in $C_{33}$, amounting to 18 % at $y_C = 0.037$.

### 4. Conclusions

The elastic properties of Fe–C and Fe–N martensites, together with those of bcc iron, have been studied by means of first-principles calculations using the supercell method. The six independent elastic constants have been determined at two solute concentrations, $y = 0.0185$ and $0.0370$. While absolute magnitudes of the equilibrium interatomic distances and the elastic constants of bcc iron are at some variance with experiment, and the elastic anisotropy of bcc iron is not satisfactorily reproduced, relative changes with the solute content, i.e., their concentration dependencies, appear reasonable in the light of the experimental data of the elastic moduli of polycrystals. All the elastic stiffness constants decrease with the solute content, roughly by 10 % at $y = 0.037$, while $C_{11}$ and $C_{33}$ decrease appreciably less and more rapidly, respectively, leading to important departure from each other. The significant decrease of $C_{33}$ must reflect weakening of atomic bonding in the direction parallel to the tetragonal axis, which could result from increased distances between the host atoms in that direction.

**Acknowledgements**

This study is supported by CREST Basic Research Program on 'Creation of innovative functions of intelligent materials on the basis of element strategy' provided by Japan Science and Technology Agency, and by cooperative research programmes of the Centre for Computational Materials Science of the Institute for Materials Research (CCMS-IMR), Tohoku University. We are indebted to Professor Furuhara T. (Tohoku University) for the leadership role and encouragement in the element strategy project on iron and steel, and to Professors Kawazoe Y., Mizuseki H., and R. Belosludov for coordinating the research using SR16000 supercomputing facilities at CCMS-IMR. We also wish to thank Professor Murata Y. (Nagoya University) for motivating this work, as well as for his continued interest and stimulating discussion, Professors M. H. F. Sluiter (Delft University of Technology) and Chen Y. (Tohoku University) for useful advices and enlightening discussion, and Dr Ohtsuka H. (National Institute for Materials Science) for informing us of the stabilities of C–C configurations prior to publication.

(4,191 words)

# REFERENCES

1) G. Krauss: Steels – Processing, structure, and performance, ASM International, Materials Park, Ohio (2005), Chap. 12.

2) D. V. Edmonds, K. He, F. C. Rizzo, B. C. De Cooman, D. K. Matlock and J. G. Speer: *Mater. Sci. Eng. A*, **438–440** (2006), 25.

3) T. Tsuchiyama, J. Tobata, T. Tao, N. Nakada and S. Takaki: *Mater. Sci. Eng. A*, **532** (2012), 585.

4) M. Doi: *Prog. Mater. Sci.*, **40** (1996), 79.

5) A. Onuki: Phase transition dynamics, Cambridge University Press, Cambridge (2002), Chap. 10.

6) Y. Tsukada, Y. Murata, T. Koyama and M. Morinaga: *Mater. Trans.*, **50** (2009), 744.

7) A. Artemev, Y. Jin and A. G. Khachaturyan: *Acta Mater.*, **49** (2001), 1165.

8) J. Kundin, D. Raabe and H. Emmerich: *J. Mech. Phys. Solids*, **59** (2011), 2082.

9) Z. Cong, Y. Murata, Y. Tsukada and T. Koyama: *Mater. Trans.*, **53** (2012), 1822.

10) T. W. Heo and L.-Q. Chen: *Acta Mater.*, **76** (2014), 68.

11) E. Schmidtmann, E. Hougardy and H. Schenck: *Arch. Eisenhüttenw.*, **36** (1965), 191.

12) B. N. Dey, J. J. Gilman and A. E. Nehrenberg: *Philos. Mag.*, **24** (1971), 1257.

13) G. R. Speich, A. J. Schwoeble and W. C. Leslie: *Metall. Trans.*, **3** (1972), 2031.

14) G. R. Speich and W. C. Leslie: *Metall. Trans.*, **4** (1973), 1873.

15) S. A. Kim and W. L. Johnson: *Mater. Sci. Eng. A*, **452–453** (2007), 633.

16) N. Gunkelmann, H. Ledbetter and H. M. Urbassek: *Acta Mater.*, **60** (2012), 4901.

17) R. F. S. Hearmon: *Acta Crystallogr.*, **6** (1953), 331.

18) M. J. Mehl, B. M. Klein, D. A. Papaconstantopoulos, Intermetallic compounds: Vol. 1 Principles, ed. by J. H. Westbrook and R. L. Fleischer, John-Wiley & Sons, Chichester (1994), Chap. 9.

19) W. Boas and J. K. Mackenzie: *Prog. Met. Phys.*, **2** (1950), 90.

20) G. Kresse and J. Furthmüller: *Comp. Mater. Sci.*, **6** (1996), 15.

21) P. E. Blöchl: *Phys. Rev. B*, **50** (1994), 17953.

22) J. P. Perdew, K. Burke and M. Ernzerhof: *Phys. Rev. Lett.*, **77** (1996), 3865.

23) H. Ohtsuka, V. A. Dinh, T. Ohno, K. Tsuzaki, K. Tsuchiya, R. Sahara, H. Kitazawa and T. Nakamura: *Tetsu-to-Hagané*, **100** (2014), 149.

24) L. Cheng, A. Böttger, Th. H. Keijser and E. J. Mittemeijer: *Scripta Metall. Mater.*, **24** (1990), 509.

25) S. S. Brenner: Growth and perfection of crystals, edi. by R. H. Doremus, B. W. Roberts and D. Turnbull, John Willey & Sons, New York (1958), 157.

26) B. E. Powell and M. J. Skove: *Phys. Rev.*, **174** (1968), 977.

27) J. A. Rayne and B. S. Chandrasekhar: *Phys. Rev.*, **122** (1961), 1714.

28) L. Vočadlo, G. A. de Wijs, G. Kresse, M. Gillan and G. D. Price: *Faraday Discuss.*, **106** (1997), 205.

29) G. Y. Guo and H. H. Wang: *Chin. J. Phys.* (Taipei), **38** (2000), 949.

30) K. J. Caspersen, A. Lew, M. Ortiz and E. A. Carter: *Phys. Rev. Lett.*, **93** (2004), 115501.

31) X. Sha and R. E. Cohen: *Phys. Rev. B*, **74** (2006), 214111.

32) H. L. Zhang, B. Johansson and L. Vitos: *Phys. Rev. B*, **79** (2009), 224201.

33) S. L. Shang, A. Saengdeejing, Z. G. Mei, D. E. Kim, H. Zhang, S. Ganeshan, Y. Wang and Z. K. Liu: *Comp. Mater. Sci.*, **48** (2010), 813.

34) F. C. Nix and D. MacNair: *Phys. Rev.*, **60** (1941), 597.

35) J. F. Nye: Physical properties of crystals – Their representation by tensors and matrices, Oxford University Press, Oxford (1957).

36) R. F. S. Hearmon: *Adv. Phys.*, **5** (1956), 323.

37) R. Hill: *Proc. Phys. Soc. A*, **65** (1952), 349.

TABLE CAPTIONS

Table 1. Strain tensors employed for computing the elastic stiffness constants of cubic crystals, and the coefficients of the second- and third-order terms in the strain energy.

Table 2. Strain tensors employed for computing the elastic stiffness constants of tetragonal crystals, and the coefficients of the second- and third-order terms in the strain energy.

Table 3. Lattice parameter $a$, the second- and third-order bulk moduli $B$ and $B^{(3)}$, the second-order elastic stiffness constants $C'$, $C_{44}$, $C_{11}$ and $C_{12}$, and the shear anisotropy factor $A = C_{44}/C'$, of bcc iron. Numbers in brackets indicate references.

Table 4. Second-order elastic stiffness constants of bct Fe–C and Fe–N alloys (in GPa). $y$ is the solute concentration.

Table 5. Third-order elastic stiffness coefficients in deformation of type 1 for bct Fe–C and Fe–N alloys (in TPa). $y$ is the solute concentration.

Table A.1. Second-order elastic compliance constants $S_{ij}$ of bcc iron and bct Fe–C and Fe–N alloys (in $10^{-12}$ Pa$^{-1}$). $y$ is the solute concentration. The lattice parameters, $a$ and $c$ (in $10^{-12}$ m), are computed from the dimensions of the fully-optimized supercell (cf. Fig. 1).

FIGURE CAPTIONS

Fig. 1. Lattice parameters and axial ratio of (a) Fe–C and (b) Fe–N alloys. The dashed lines are experimental data at room temperature assessed by Cheng et al.[24].

Fig. 2. Strain energy plotted against the square of the strain parameter, $x$, in bcc iron. The down and up triangles for deformation 1 are the results from negative and positive values of $x$, respectively. The solid lines for 2a and 2b are fitted to the selected data points on which the lines are drawn.

Fig. 3. Strain energy as a function of strain, $x$, in deformation 1 of bcc iron. The solid curve is a cubic polynomial fitted to the selected data points indicated by the vertical markers at the bottom.

Fig. 4. Strain energy plotted against the square of the strain parameter, $x$, for deformation 1a, 1b, and 1c of bct $Fe_{54}C_1$ alloy. The down and up triangles indicate the results from negative and positive values of $x$, respectively.

Fig. 5. Strain energy as a function of strain parameter, $x$, for deformation 1a, 1b, and 1c of bct $Fe_{54}C_1$ alloy. The solid curves are cubic polynomials fitted to the selected data points indicated by the vertical markers at the bottom.

Fig. 6. Strain energy plotted against the square of the strain parameter, $x$, for deformation 2a, 2b, and 2c of bct $Fe_{54}C_1$ alloy. The solid lines are linear functions fitted to the selected data points on which the lines are drawn.

Fig. 7.  Elastic stiffness constants as a function of solute concentration $y$ for (a) Fe–C and (b) Fe–N alloys. The experimental values of bcc iron at 4.2 K[27] are indicated by horizontal arrows. The elastic constants of iron and Fe–C martensite reported by Kim and Johnson[15] are shown in (a) (in green, symbols connected by dotted lines).

Fig. 8.  3D surface plots of Young's modulus $E$ (in GPa) of (a) bcc iron at 4.2 K[27], (b) bcc iron, (c) bct Fe–C alloy of $y_C = 0.037$, and (d) bct Fe–N alloy of $y_N = 0.037$.

Fig. 9.  (a) (001), (b) (100) and (c) ($1\bar{1}0$) cross-sectional views of Young's modulus $E$ of bcc iron at 4.2 K[27] ('expt', dashed curve), and those calculated for bcc iron and bct Fe–C and Fe–N alloys. The axis scale is in GPa.

Fig. 10.  3D surface plots of rigidity modulus $G$ (in GPa) of (a) bcc iron at 4.2 K[27], (b) bcc iron, (c) bct Fe–C alloy of $y_C = 0.037$, and (d) bct Fe–N alloy of $y_N = 0.037$.

Fig. 11.  (a) (001), (b) (100) and (c) ($1\bar{1}0$) cross-sectional views of the rigidity modulus $G$ of bcc iron at 4.2 K[27] ('expt', dashed curve), and those calculated for bcc iron and bct Fe–C and Fe–N alloys. The axis scale is in GPa.

Fig. 12.  Rigidity modulus $G$, Young's modulus $E$, bulk modulus $B$, and Poisson's ratio $\nu$ of polycrystalline aggregates of (a) Fe–C alloys and (b) Fe–N alloys calculated from the single-crystal elastic constants by the methods of Voigt, Reuss, and Voigt-Reuss-Hill.